\documentclass{article}

 \usepackage[dblblindworkshop, final]{neurips_2025}
\workshoptitle{Lock-LLM}

\usepackage[utf8]{inputenc} 
\usepackage[T1]{fontenc}    
\usepackage{hyperref}       
\usepackage{url}            
\usepackage{booktabs}       
\usepackage{amsfonts}       
\usepackage{nicefrac}       
\usepackage{microtype}      
\usepackage{xcolor}         
\usepackage{colortbl}
\usepackage{array}
\usepackage{multirow}
\usepackage[table]{xcolor}
\usepackage{caption}
\usepackage{amsmath}
\usepackage{graphicx}
\usepackage{wrapfig}

\title{DistilLock: Safeguarding LLMs from Unauthorized Knowledge Distillation on the Edge}

\author{%
Asmita Mohanty\textsuperscript{1}, 
Gezheng Kang\textsuperscript{2}, 
Lei Gao\textsuperscript{1}, 
Murali Annavaram\textsuperscript{1} \\
\textsuperscript{1}University of Southern California, 
\textsuperscript{2}University of California, Davis \\
\textsuperscript{1}\texttt{\{asmitamo, leig, annavara\}@usc.edu}, 
\textsuperscript{2}\texttt{gzkang@ucdavis.edu}
}
\begin{document}

\maketitle

\begin{abstract}

Large Language Models (LLMs) have demonstrated strong performance across diverse tasks, but fine-tuning them typically relies on cloud-based, centralized infrastructures. This requires data owners to upload potentially sensitive data to external servers, raising serious privacy concerns. An alternative approach is to fine-tune LLMs directly on edge devices using local data; however, this introduces a new challenge: the model owner must transfer proprietary models to the edge, which risks intellectual property (IP) leakage.
To address this dilemma, we propose DistilLock, a TEE-assisted fine-tuning framework that enables privacy-preserving knowledge distillation on the edge. In DistilLock, a proprietary foundation model is executed within a trusted execution environment (TEE) enclave on the data owner’s device, acting as a secure black-box teacher. This setup preserves both data privacy and model IP by preventing direct access to model internals.
Furthermore, DistilLock employs a model obfuscation mechanism to offload obfuscated weights to untrusted accelerators for efficient knowledge distillation without compromising security. We demonstrate that DistilLock prevents unauthorized knowledge distillation processes and model-stealing attacks while maintaining high computational efficiency, but offering a secure and practical solution for edge-based LLM personalization. 

\end{abstract}

\section{Introduction}
Proprietary Large Language Models (LLMs) such as GPT~\cite{openai2022chatgpt}, Gemini~\cite{google2023gemini}, and Claude~\cite{anthropic2024claude} have demonstrated superior performance compared to open-source models, largely due to the availability of extensive computing infrastructure, high-quality datasets, and advanced optimization pipelines. As a result, their trained weights are considered valuable intellectual property and are typically not released to the public. Downstream users who wish to customize these models must rely on cloud-based APIs and submit labeled data to remote servers for fine-tuning. This centralized setup not only incurs cloud costs but also raises serious privacy concerns, particularly for sensitive or proprietary datasets.

The growing availability of edge-centric ML accelerators~\cite{qualcomm_hexagon, nvidia_jetson} and resource-efficient training algorithms~\cite{zhu2023pockengine, gao2024mobizo} has made it feasible to fine-tune LLMs with billions of parameters directly on edge devices. This shift offers a compelling privacy advantage, as user data remains local and never leaves the device. However, this alternative comes with a new risk: the model owner must send their proprietary foundation model to an untrusted edge environment, thereby exposing sensitive model internals and violating intellectual property boundaries.

Recent works have explored methods to mitigate this dilemma by enabling partial decentralization of the fine-tuning process. For instance, Offsite-Tuning~\cite{xiao2023offsitetuning} compresses the foundation model and deploys it on the user’s device as a lightweight backbone to train local adapters, which are then sent back to the model owner for integration. Split-and-Privatize~\cite{shen2023splitandprivatize} divides the model into two parts, training the first half locally and transmitting privatized intermediate embeddings to the server, where the second half completes the forward and backward pass. While these solutions reduce data exposure, they sacrifice training performance due to model compression or differential privacy, and still depend on the remote model owner to perform inference after training.

To overcome these limitations, we introduce DistilLock, a secure on-device knowledge distillation framework that protects both user data privacy and foundation model confidentiality. DistilLock uses Trusted Execution Environments (TEEs) to run the proprietary foundation model as a black-box teacher inside a hardware-protected enclave on the user's device. This allows knowledge distillation to happen entirely on-device without exposing the model weights or user data. Crucially, to avoid the computational overhead of a full model forward pass within the TEE, DistilLock employs a model obfuscation strategy that enables most compute-heavy operations to be offloaded to untrusted accelerators (e.g., GPUs). The TEE is used only for lightweight authorization, and the offloaded weights are obfuscated so they remain functional but cannot be reverse-engineered to reveal sensitive model information. Once distillation completes, the obfuscated teacher is removed, leaving only the distilled student model tailored to the user’s task.

We evaluate DistilLock across multiple foundation models and diverse downstream tasks. Our results show that DistilLock prevents unauthorized knowledge distillation and resists model-stealing attacks, even when adversaries attempt surrogate training on the obfuscated model. Furthermore, we analyze the efficiency of our framework and demonstrate that DistilLock introduces only minimal computation overhead through lightweight TEE authorization and obfuscated model offloading.

\section{Background and Threat Model}
\label{sec:background}
\textbf{Trusted Execution Environment.} 
Trusted Execution Environments (TEEs) are hardware-enforced secure regions that provide confidentiality and integrity guarantees for code and data during execution. Widely used implementations include Intel SGX~\cite{mckeen2013innovative}, Intel TDX~\cite{cheng2023inteltdxdemystifiedtopdown}, and ARM TrustZone~\cite{alves2004trustzone}. Despite architectural differences, they share the common goal of shielding sensitive computation from untrusted system software such as the OS or hypervisor. In this work, we prototype our system using Intel SGX, which offers fine-grained enclave abstraction for user-space applications and is readily available on edge CPUs. Although recent GPUs like the NVIDIA H100 support TEEs~\cite{nvidia_h100}, such capabilities are not yet widely deployed in the edge environments we target.

\textbf{Knowledge Distillation.}  
Knowledge distillation (KD) transfers knowledge from a large, high-capacity teacher model to a smaller student model by training the student to mimic the teacher's behavior. The student learns from both the ground-truth labels and the teacher’s soft output distribution, enabling it to achieve comparable performance with significantly fewer resources.
Let $y$ be the ground-truth label, $q_S$ and $q_T$ be the logits from the student and teacher, respectively, and let $q^{(\tau)} = \text{softmax}(q / \tau)$ denote the softened output with temperature $\tau > 1$. The distillation loss is defined as:
$
\mathcal{L}_{\text{KD}} = \alpha \cdot \mathcal{L}_{\text{CE}}(y, q_S) + \beta \cdot \mathcal{L}_{\text{KL}}(q_T^{(\tau)}, q_S^{(\tau)}),
$
where $\mathcal{L}_{\text{CE}}$ is the standard cross-entropy loss and $\mathcal{L}_{\text{KL}}$ is the Kullback--Leibler divergence between the teacher and student distributions. 
While many advanced variants of the KD algorithm have been proposed~\cite{gu2024minillm, DISTILLM, agarwal2024onpolicy}, we focus on the vanilla formulation in this work, as our primary goal is to study the security and system-level aspects of on-device distillation.

\textbf{Threat Model.}
We define two primary parties: the defender and the attacker. The defender owns the model deployed on an edge device, while the attacker's goal is to steal it.

\textit{Defender’s Goal:}
The defender's primary objective is to ensure their deployed teacher model, $\mathcal{M}_T$, generates the correct logits, $q_T$, only when proper authorization is given by a TEE. To maintain efficiency, the defender offloads the majority of the computation to an untrusted GPU, which can be accessed in a white-box manner by the user. The defender's ultimate goal is to degrade white-box attacks to a black-box setting, preventing the attacker from accessing the model's weights and intellectual property.

\textit{Adversary’s Goal and Capability:}
The attacker's goal is to develop an independent surrogate model, $\mathcal{M}_T'$, that can replicate the performance of the authorized teacher model, $\mathcal{M}_T$. Our threat model assumes that the attacker has white-box access to the model details, including architecture and weights, for any part of the model that resides outside the TEE (e.g., on the GPU). The attacker can use existing techniques \cite{chen2022copy, chen2022teacher} to infer the full architecture and obtain the weights of these exposed components. We also assume the attacker possesses a well-labeled dataset for the targeted task, which they can use to facilitate the attack. The TEE itself, however, is considered a secure and uncompromisable environment.

\section{DistilLock's Defense Mechanism}

TEEs often lack support for powerful accelerators like GPUs, making them approximately 50 times slower for running LLMs \cite{Li_2024}. Therefore, different methods \cite{elgamal2020serdab, mo2020darknetz, sun2023shadownet} have been proposed to shield and execute partial models inside TEEs and offload the remaining parts to an untrusted environment. Inspired by recent works \cite{Li_2024, yuan2024securetransformerinferenceprotocol}, we leverage model obfuscation to ensure model and data confidentiality while preserving utility and efficiency. DistilLock consists of two steps: model obfuscation and model authorization. 

\textbf{Model Obfuscation.} Let $x \in \mathbb{R}^{n \times d}$ denote the input where $n$ is the sequence length (e.g., the number of tokens) and $d$ is the model dimension. We define the forward pass in a Transformer block in typical LLMs like Llama \cite{grattafiori2024llama3herdmodels} and Qwen \cite{bai2023qwentechnicalreport} as follows:

\begin{equation}
\begin{aligned}
Q &= xW_q, \quad K = xW_k, \quad V = xW_v, &\quad W_q, W_k, W_v &\in \mathbb{R}^{d \times d}, \\
u &= \text{Softmax} \left( \frac{QK^T}{\sqrt{k}} \right) VW_o, &\quad W_o &\in \mathbb{R}^{d \times d}, \\
v &= \text{LayerNorm}(u + x; \gamma_1, \beta_1), &\quad \gamma_1, \beta_1 &\in \mathbb{R}^{d}, \\
z &= (\text{SiLU}(vW_1)vW_3)W_2, &\quad W_1, W_3 \in \mathbb{R}^{d \times m}, \quad W_2 &\in \mathbb{R}^{m \times d}, \\
y &= \text{LayerNorm}(z + v; \gamma_2, \beta_2), &\quad \gamma_2, \beta_2 &\in \mathbb{R}^{d},
\end{aligned}
\end{equation}

Let $\pi_{emb} \in \{0,1\}^{vocab \times vocab}$, $\pi \in \{0,1\}^{d \times d}$ denote two permutation matrices. Before the model is transferred to the user side, the model owner first transforms the parameters as follows:
\begin{equation}
\begin{aligned}
W_{emb}' = \pi_{emb}^T W_{emb}, \\
W_q' = \pi^T W_q, \quad W_k' = \pi^T W_k, \quad W_v' = \pi^T W_v, \quad  W_o' = W_o \pi, \\
W_1' = \pi^T W_1, \quad W_3' = \pi^T W_3, \quad W_2' = W_2 \pi,  \\
\quad \gamma_1' = \gamma_1 \pi, \quad \beta_1' = \beta_1 \pi, \quad \gamma_2' = \gamma_2 \pi, \quad \beta_2' = \beta_2 \pi,  \\
W_{cls}' = \pi^T W_{cls},
\end{aligned}
\end{equation}
where $W_{emb} \in \mathbb{R}^{vocab \times d}$ is the word embedding layer and $W_{cls} \in \mathbb{R}^{d \times vocab}$ is the final linear classifier. 

\begin{figure}
    \centering
    \includegraphics[width=1\linewidth]{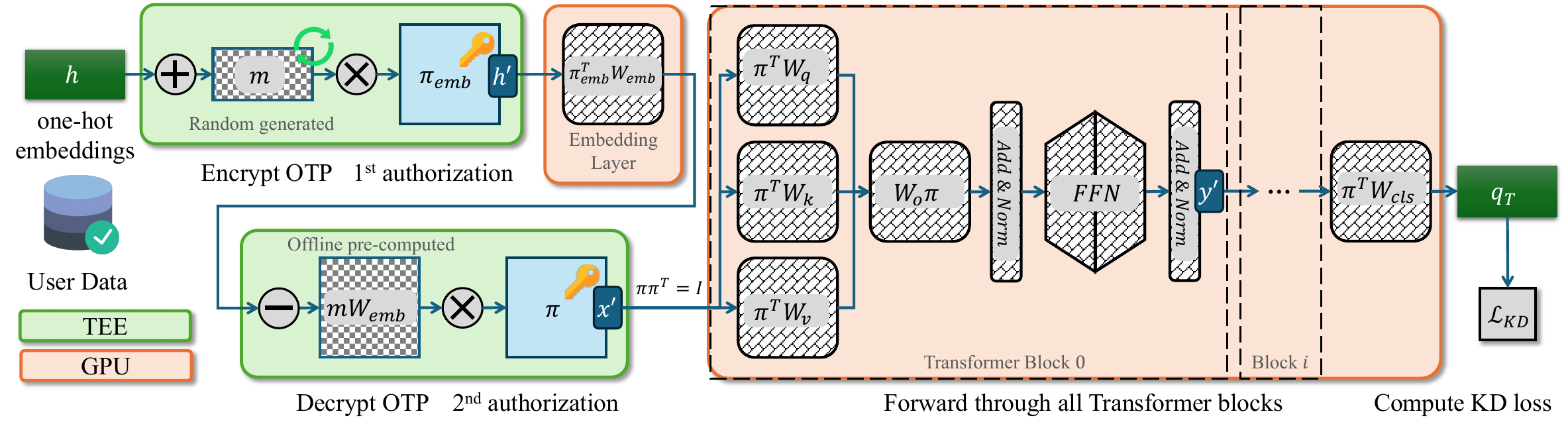}
    \caption{Overview of DistilLock. The proprietary teacher model is obfuscated before deployment and executed with TEE authorization on the user’s device. The TEE applies the correct permutation and one-time pad to the input, while compute-intensive operations run on untrusted accelerators using obfuscated weights. This design preserves model functionality for authorized knowledge distillation while preventing unauthorized access or model extraction.}
    \label{fig:method}
    \vspace{-4pt}
\end{figure}

\textbf{Model Authorization.} 
The obfuscated model and the encrypted permutation matrix $\pi$ are delivered to the user, who then instantiates the TEE. As shown in Figure~\ref{fig:method}, the user’s input $h \in \mathbb{R}^{n \times vocab}$ (represented as one-hot vectors) is first sent to the TEE. Inside the TEE, $h$ is encrypted with a one-time pad (OTP) $m \in \mathbb{R}^{n \times vocab}$ and multiplied by $\pi_{emb}$ on the GPU: $h' = (h + m)\pi_{emb}.$ This operation hides both $m$ and $\pi_{emb}$ from the user. The encrypted and permuted input $h'$ then exits the TEE and is multiplied by the obfuscated embedding layer: $h' W'_{emb} = (h + m)\pi_{emb} \cdot \pi_{emb}^{T} W_{emb} = (h + m)W_{emb},$ since $\pi_{emb}\pi_{emb}^{T} = I$. The result $(h + m)W_{emb}$ is returned to the TEE, where the OTP is removed and a new permutation is applied: $x' = \big((h + m)W_{emb} - mW_{emb}\big)\pi = hW_{emb}\pi = x\pi,$ with $mW_{emb}$ pre-computed offline.

The functionality of each subsequent Transformer block is preserved with the obfuscated parameters.
\begin{equation}
\begin{aligned}
Q' = x' W_q' = x\pi \pi^T W_q = Q, \quad K' = x' W_k' = x\pi \pi^T W_k = K, \quad V' = x' W_v' = x\pi \pi^T W_v = V, \\
u' = \text{Softmax} \left( \frac{QK^T}{\sqrt{k}} \right) V'W_o'  = \text{Softmax} \left( \frac{QK^T}{\sqrt{k}}\right) VW_o \pi = u \pi, \\
v' = \text{LayerNorm}(u' + x\pi; \gamma_1\pi, \beta_1\pi) = \text{LayerNorm}(u\pi + x\pi; \gamma_1\pi, \beta_1\pi) = v \pi, \\
z' = (\text{SiLU}(v'W_1')v'W_3')W2' = (\text{SiLU}(v\pi \pi^TW_1)v\pi \pi^TW_3)W_2\pi=z\pi, \\
y' = \text{LayerNorm}(z' + v'; \gamma_2\pi, \beta_2\pi) = \text{LayerNorm}(z \pi + v \pi; \gamma_2 \pi, \beta_2 \pi) = y \pi,
\end{aligned}
\end{equation}

The output $y'$ of this Transformer block will be the input of the next Transformer block. At the last Transformer block, the output $y'_{last}$ is multipled with the final linear classifier $W_{cls}'$ to produce the correct logits $q_T=y'_{last}W_{cls}'=y_{last} \pi \pi^TW_{cls}$.
Since the permutation matrix $\pi$ is maintained throughout the output of each decoder layer, encrypting the input for the subsequent layers, TEE computation is only required for the initial word embedding layer. Without the authorization in TEE by applying the correct $\pi_{emb}$ and $\pi$ to input, the teacher model will not produce the correct logits $q_T$, thereby locking the KD process. All other operations, including loss evaluation and the student model’s forward and backward passes, proceed as in standard KD on GPU.

\textbf{Security Analysis. }
Our security analysis focuses on a model-stealing threat model, where an adversary attempts to extract model functionality by exploiting query responses and performing surrogate training.
An attacker might attempt to recover the original parameters by guessing the permutation matrix $\pi$. However, the probability of guessing the correct matrix is approximately $\frac{1}{d!}$, where $d$ is the dimension of the model. This is computationally infeasible for foundational LLMs.

In addition, since the regular word embedding layer is a look-up operation with negligible compute cost, while in DistilLock, it becomes a matrix multiplication operation after adding one-time pad noise (i.e., $(h+m)W_{emb}$), which could add additional overhead. We limit the one-time pad noise $m$ to be a random spare k-hot vector with a number of hot keys $k \geq\frac{h}{log_2(h)}$, which reduces the expensive matrix multiplication operation to a vector summation operation while still having comparable entropy to a uniform random vector to preserve the security guarantees.

\section{Experiments}

\subsection{Distillation Lockdown}

\begin{wrapfigure}{r}{0.5\linewidth}
\vspace{-5em}
    \centering
    \includegraphics[width=\linewidth]{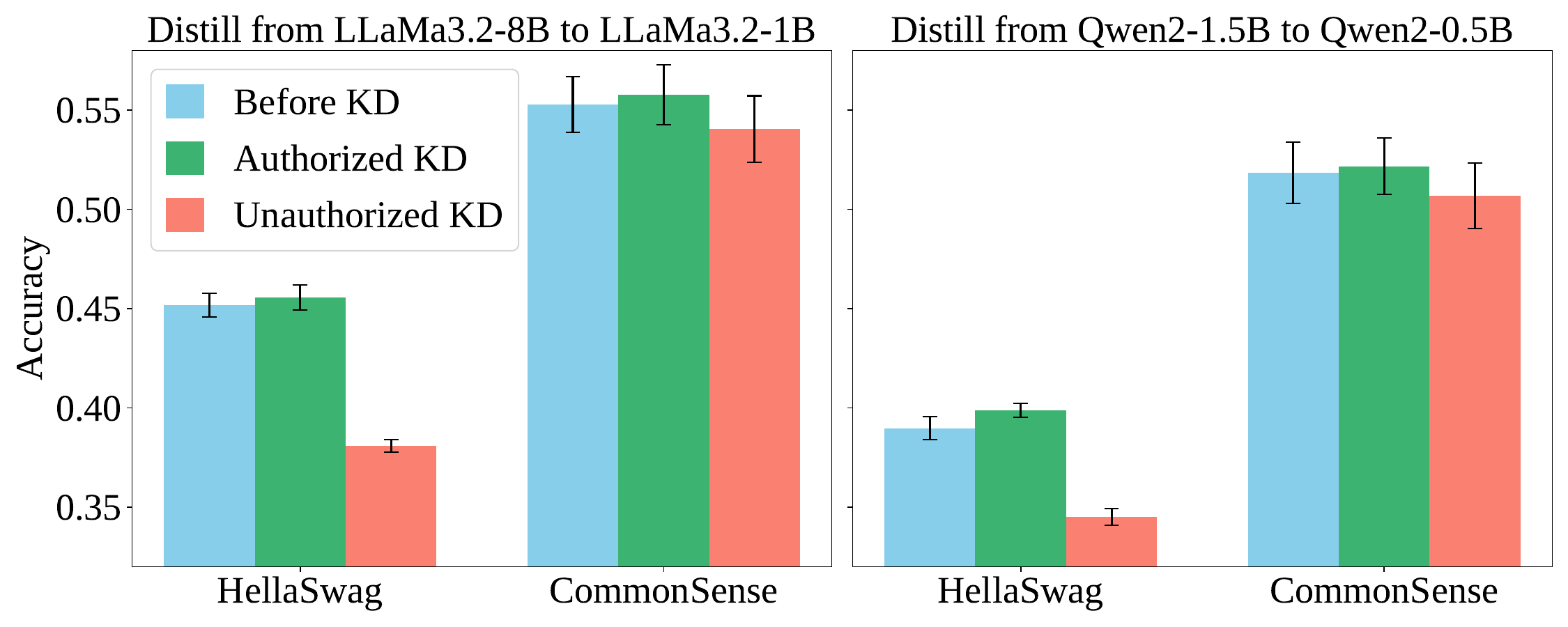}
    \caption{Knowledge distillation results under DistilLock. Authorized distillation improves student accuracy over pre-distillation baselines, while unauthorized distillation collapses performance due to random logits from the obfuscated teacher.}
    \label{fig:kd-lockdown}
\end{wrapfigure}

We evaluate DistilLock under two scenarios: (i) the user is authorized to perform knowledge distillation on their local device after receiving the obfuscated teacher model from the model owner, and (ii) the user is unauthorized, where the input embeddings lack the correct transformation $\pi_{emb}$ and $\pi$ before being passed through the teacher model. We use LLaMA3.1-8B and Qwen2-1.5B as teacher models, with LLaMA3.2-1B and Qwen2-0.5B serving as the corresponding student models \cite{grattafiori2024llama3herdmodels, bai2023qwentechnicalreport}. The student models are fine-tuned on the Alpaca dataset \cite{alpaca} under teacher guidance and evaluated on the HellaSwag \cite{zellers2019hellaswagmachinereallyfinish} and CommonsenseQA \cite{talmor-etal-2019-commonsenseqa} benchmarks. As shown in Figure~\ref{fig:kd-lockdown}, authorized knowledge distillation yields accuracy gains over pre-distillation baselines, whereas unauthorized distillation under DistilLock collapses student performance because the obfuscated teacher produces random logits that provide no useful learning signal.

\definecolor{lowest}{RGB}{255,218,185}
\definecolor{ours}{RGB}{144,238,144}

\begin{table}[htbp]
\centering
\caption{Evaluation of surrogate model attack accuracy under different defense schemes. The last row reports average accuracy relative to the black-box baseline. The lowest attack accuracy for each task is shown in green, and the second-lowest in red. DistilLock consistently achieves the lowest attack accuracy across tasks, demonstrating effectiveness in mitigating surrogate model extraction.}
\scriptsize
\begin{tabular}{cl*{7}{c}} 
\toprule
Models & Tasks  & BlackBox & WhiteBox & DarkneTZ & ShadowNet & Serdab & \textbf{DistilLock} \\
\midrule
\multirow{8}{*}{\textbf{LLaMA3.2-3B}}
& SST-2 & 67.43\% & 92.43\% & 92.54\% & 92.55\% & \cellcolor{lowest}67.51\% & \cellcolor{ours}63.17\% \\
& ARC-e & 29.14\% & 74.31\% & 30.84\% & 73.95\% & \cellcolor{lowest}29.68\% & \cellcolor{ours}25.32\% \\
& ARC-c & 29.93\% & 56.80\% & \cellcolor{lowest}30.41\% & 48.98\% & 30.53\% & \cellcolor{ours}28.53\% \\
& BoolQ & 64.20\% & 78.30\% & 78.40\% & 80.40\% & \cellcolor{lowest}65.94\% & \cellcolor{ours}65.00\% \\
& HellaSwag & 26.30\% & 57.30\% & 54.90\% & 37.10\% & \cellcolor{lowest}26.30\% & \cellcolor{ours}24.90\% \\
& RTE & 53.79\% & 79.78\% & 80.14\% & 74.01\% & \cellcolor{lowest}62.11\% & \cellcolor{ours}54.51\% \\
& QQP & 66.60\% & 80.80\% & 81.60\% & 78.80\% & \cellcolor{lowest}68.50\% & \cellcolor{ours}66.70\% \\
& QNLI & 54.90\% & 82.80\% & 79.10\% & 82.40\% & \cellcolor{lowest}67.10\% & \cellcolor{ours}56.10\% \\
\midrule
\multirow{8}{*}{\textbf{Qwen2.5-3B}}
& SST-2 & 66.63\% & 93.69\% & 91.28\% & 93.23\% & \cellcolor{lowest}69.71\% & \cellcolor{ours}65.55\% \\
& ARC-e & 27.69\% & 88.52\% & 28.41\% & 83.61\% & \cellcolor{lowest}28.05\% & \cellcolor{ours}25.87\% \\
& ARC-c & 26.53\% & 77.89\% & \cellcolor{lowest}26.53\% & 71.43\% & 29.53\% & \cellcolor{ours}24.49\% \\
& BoolQ & 52.50\% & 82.40\% & 78.80\% & 80.80\% & \cellcolor{lowest}62.50\% & \cellcolor{ours}51.30\% \\
& HellaSwag & 26.30\% & 70.40\% & 68.90\% & 68.90\% & \cellcolor{lowest}26.90\% & \cellcolor{ours}25.10\% \\
& RTE & 45.85\% & 83.03\% & 84.84\% & 83.03\% & \cellcolor{lowest}51.26\% & \cellcolor{ours}48.74\% \\
& QQP & 71.20\% & 84.30\% & 80.90\% & 83.80\% & \cellcolor{lowest}69.81\% & \cellcolor{ours}68.70\% \\
& QNLI & 53.90\% & 85.80\% & 90.10\% &\cellcolor{lowest} 51.00\% & 61.40\% & \cellcolor{ours}56.60\% \\
\midrule
\multirow{8}{*}{\textbf{Qwen2.5-1.5B}}
& SST-2 & 66.62\% & 93.58\% & 84.73\% & 93.35\% & \cellcolor{lowest}69.74\% & \cellcolor{ours}65.68\% \\
& ARC-e & 30.96\% & 87.25\% & \cellcolor{lowest}27.76\% & 79.31\% & 28.23\% & \cellcolor{ours}27.51\% \\
& ARC-c & 28.57\% & 79.93\% & \cellcolor{lowest}29.36\% & 64.29\% & 29.93\% & \cellcolor{ours}27.64\% \\
& BoolQ & 52.70\% & 79.40\% & 76.70\% & 79.50\% &\cellcolor{lowest} 69.60\% & \cellcolor{ours}52.50\% \\
& HellaSwag & 25.70\% & 60.60\% & 63.23\% & 60.70\% & \cellcolor{lowest}27.30\% & \cellcolor{ours}25.30\% \\
& RTE & 46.57\% & 77.98\% & 79.78\% & 68.95\% & \cellcolor{lowest}54.87\% & \cellcolor{ours}46.43\% \\
& QQP & 71.80\% & 84.10\% & 83.10\% & 84.40\% & \cellcolor{lowest}74.60\% & \cellcolor{ours}70.40\% \\
& QNLI & 52.80\% & 85.30\% & 86.00\% & 82.10\% & \cellcolor{lowest}62.30\% & \cellcolor{ours}52.40\% \\
\midrule
\multirow{8}{*}{\textbf{Qwen2-1.5B}}
& SST-2 & 66.63\% & 93.23\% & 84.12\% & 91.40\% & \cellcolor{lowest}70.98\% & \cellcolor{ours}67.09\% \\
& ARC-e & 30.96\% & 78.69\% & \cellcolor{lowest}28.31\% & 77.76\% & 28.68\% & \cellcolor{ours}25.13\% \\
& ARC-c & 28.91\% & 79.10\% & \cellcolor{lowest}26.12\% & 54.42\% & 29.77\% & \cellcolor{ours}27.93\% \\
& BoolQ & 52.70\% & 81.50\% & 76.70\% & 76.50\% &\cellcolor{lowest} 61.30\% & \cellcolor{ours}51.70\% \\
& HellaSwag & 26.30\% & 59.70\% & 63.74\% & 54.90\% & \cellcolor{lowest}26.53\% & \cellcolor{ours}24.10\% \\
& RTE & 46.57\% & 83.03\% & 78.34\% & 76.53\% &\cellcolor{lowest} 50.18\% & \cellcolor{ours}48.01\% \\
& QQP & 70.40\% & 84.30\% & 81.70\% & 84.70\% & \cellcolor{lowest}74.80\% & \cellcolor{ours}70.40\% \\
& QNLI & 52.80\% & 85.80\% & 84.50\% & 74.10\% & \cellcolor{lowest}66.90\% & \cellcolor{ours}52.45\% \\
\midrule
\multicolumn{2}{c}{\textbf{Relative Average Attack Accuracy}} & 1.00$\times$ & 1.69$\times$ & 1.41$\times$ & 1.58$\times$ & 1.09$\times$ & 0.98$\times$ \\
\bottomrule
\end{tabular}
\label{tab:adversarial_training}
\end{table}

\subsection{Adversarial Training}

While adversarial users can be prevented from performing knowledge distillation, they may still attempt to extract the teacher model by training a surrogate that replicates its performance, as described in Section~\ref{sec:background}. To evaluate the robustness of defense mechanisms against such model extraction attacks, we simulate surrogate training across diverse NLP tasks: AI2 Reasoning Challenge (ARC-e, ARC-c) \cite{clark2018thinksolvedquestionanswering} for science reasoning, GLUE \cite{wang2018glue} tasks including SST-2 for sentiment classification, RTE \cite{poliak2020surveyrecognizingtextualentailment} for textual entailment, QNLI \cite{demszky2018transformingquestionansweringdatasets} for question-answering, and QQP \cite{sharma2019naturallanguageunderstandingquora} for question-pair similarity, as well as HellaSwag \cite{zellers2019hellaswagmachinereallyfinish} for commonsense reasoning and BoolQ \cite{clark2019boolqexploringsurprisingdifficulty} for yes/no reading comprehension. 

We consider six attack scenarios with different adversary capabilities to comprehensively measure defense effectiveness. The white-box attack assumes full access to model weights, yielding the best achievable attack accuracy baseline, while the black-box attack assumes knowledge of the architecture only with random weight initialization, representing the worst-case baseline. Intermediate defenses include Serdab \cite{elgamal2020serdab}, which protects only the first transformer layer inside the TEE; DarkneTZ \cite{mo2020darknetz}, which shields the final transformer layer and subsequent components; and ShadowNet \cite{sun2023shadownet}, which obfuscates linear layers via matrix transformations while executing all other layers outside the TEE. Finally, we evaluate DistilLock, in which the attacker initializes the surrogate model directly with the obfuscated weights. A more detailed experimental setup is provided in Appendix \ref{sec:app-exp}.

Results in Table~\ref{tab:adversarial_training} demonstrate that DistilLock consistently achieves the lowest attack success across all evaluated tasks and models. On average, DistilLock reduces surrogate model accuracy to 0.98$\times$ relative to the black-box baseline, effectively neutralizing the attacker’s advantage. In contrast, prior defenses such as ShadowNet (1.58$\times$) and DarkneTZ (1.41$\times$) leave significant headroom for the adversary, highlighting the stronger robustness of our approach.

\subsection{Efficiency Cost}
We evaluate efficiency by measuring the fraction of total FLOPs that must be executed inside the TEE under different defense schemes. As shown in Table~\ref{tab:efficiency}, DistilLock incurs the lowest overhead, requiring less than 1.2\% of total FLOPs, compared to 3–16\% for DarkneTZ, ShadowNet, and Serdab. The efficiency gain comes from reducing permutation matrix multiplications to simple column shuffling and pre-computing the OTP decryption factor offline. These results show that DistilLock delivers stronger protection while imposing significantly lower efficiency costs.

\begin{table}[htbp]
\centering
\caption{TEE computation overhead across defense schemes and models. DistilLock requires the lowest FLOPs inside the TEE.}
\scriptsize
\begin{tabular}{clrrrr}
\toprule
Models & Total FLOPs & DarkneTZ-TEE & ShadowNet-TEE & Serdab-TEE & \textbf{DistilLock-TEE}  \\
\midrule
\textbf{LLaMA3.2-3B} & 11,913,901,394,688 & 427,016,202,048 & 1,469,885,448,192 & 425,440,192,320 & 52,539,949,056  \\
& \% of total & 3.58\% & 12.34\% & 3.57\% & 0.44\%  \\
\midrule
\textbf{Qwen2.5-1.5B} & 5,613,429,499,648 & 199,157,061,696 & 920,602,437,942 & 260,459,698,240 & 62,236,131,328  \\
& \% of total & 3.55\% & 16.40\% & 4.64\% & 1.11\%  \\
\bottomrule
\end{tabular}
\label{tab:efficiency}
\end{table}

\section{Related Work}
\textbf{TEE-assisted Model Execution.}
Serdab \cite{elgamal2020serdab} protects shallow transformer layers, which are considered more critical. ShadowNet \cite{sun2023shadownet} obfuscates all linear layers via matrix transformations and offloads them, along with the remaining layers, to untrusted hardware, but remains vulnerable to model-stealing attacks \cite{zhang2023no}. DarkneTZ \cite{mo2020darknetz} secures the final transformer layers and the output classifier inside the TEE. TransLinkGuard \cite{Li_2024} applies a permutation-based obfuscation strategy to protect model weights, but requires TEE authorization at each transformer block, introducing substantial communication overhead. In contrast, our method requires only a single TEE authorization at the beginning of the forward pass. More importantly, these methods focus on protecting models during inference, whereas our work targets protection during the fine-tuning process.

\textbf{Private Knowledge Distillation.}
RONA \cite{rona} assumes the provider has trained a foundation model on both sensitive and public data. Distillation is performed only on public data to train a compact student model whose representations approximate those of the foundation model. The student is deployed to users, while the provider retains both the foundation model and the sensitive data. To ensure formal guarantees, RONA perturbs the distilled knowledge to satisfy differential privacy. Swing Distillation \cite{swingdistillation} assumes disjoint teacher and student datasets and introduces dynamic temperature and soft target protection to reduce leakage of private information. However, these approaches primarily protect the provider’s sensitive training data in the cloud, whereas our solution focuses on protecting user data by performing fine-tuning directly on local devices.

\textbf{Efficient On-device Fine-tuning.}
Offsite-Tuning \cite{xiao2023offsitetuning} enables lightweight adapter fine-tuning on downstream tasks, assisted by a lossy compressed emulator sent from the provider to the user. The fine-tuned adapter is then returned and integrated into the full model on the provider’s side. This workflow, however, exposes the user’s fine-tuned adapters to the provider and lacks formal privacy guarantees. PockEngine \cite{zhu2023pockengine} and MobiZO \cite{gao2024mobizo} focus on memory-efficient on-device training. PockEngine employs a first-order optimizer but restricts updates to a subset of layers to reduce activation storage. MobiZO instead uses a zeroth-order optimizer, eliminating backward propagation and relying solely on inference engines to estimate gradients. However, neither approach addresses model ownership concerns, which our method explicitly considers.

\section{Conclusion}
We introduced DistilLock, a TEE-assisted fine-tuning framework that safeguards both user data privacy and model intellectual property during on-device knowledge distillation. By combining lightweight TEE authorization with model obfuscation, DistilLock prevents unauthorized distillation and model-stealing attacks while keeping computational overhead minimal. Our experiments demonstrate that DistilLock achieves secure and efficient edge-based LLM personalization, outperforming prior defenses in both robustness and efficiency.

\section*{Acknowledgment}
We sincerely thank all the reviewers for their time and constructive comments. This material is based upon work supported by NSF award number 2224319, REAL@USC-Meta center, and VMware gift. The views, opinions, and/or findings expressed are those of the author(s) and should not be interpreted as representing the official views or policies of the U.S. Government.

\bibliographystyle{unsrt}
\bibliography{refs}

\newpage
\appendix

\section{Limitations}
While DistilLock demonstrates strong protection against unauthorized knowledge distillation and model-stealing attacks, several limitations remain.  
 
First, our current implementation is built on Intel SGX and has not been extended to other trusted execution environments such as ARM TrustZone or emerging GPU TEEs. Second, although obfuscation reduces FLOPs by simplifying permutation operations, it still introduces extra steps such as OTP application and permutation management, and future adaptive attacks (e.g., side-channel leakage or advanced surrogate training) may weaken its guarantees. Finally, our evaluation is limited to standard KD settings and datasets, leaving the effectiveness of DistilLock under advanced distillation variants and more complex scenarios as an open question.


\section{LayerNorm and RMSNorm Proof}

\textbf{LayerNorm.} Layer Normalization normalizes activations across the feature dimension of each sample to stabilize training and reduce covariate shift. It has been widely adopted in early transformer architectures such as BERT and GPT-2.  

\textbf{RMSNorm.} Recent LLMs such as LLaMA and Qwen adopt Root Mean Square Normalization (RMSNorm), a simplified alternative to LayerNorm. RMSNorm removes mean-centering since the residual connection dominates, and instead normalizes only by the root mean square of the activations. This reduces computation overhead while maintaining stability.  

In both LayerNorm and RMSNorm, the statistics (mean, variance, or RMS) are permutation-invariant, but the feature-wise scaling and shifting parameters must be permuted consistently with the activations. Otherwise, functional equivalence is broken. We show that both normalization layers are \textit{permutation-equivariant} when the inputs and their parameters $(\gamma, \beta)$ are permuted by the same permutation matrix $\pi$.  

\textit{Proof of LayerNorm.}  
We aim to prove that  
$$
\text{LayerNorm}(x\pi;\gamma\pi,\beta\pi) = \text{LayerNorm}(x;\gamma,\beta)\pi.
$$

The LayerNorm function is defined for $x \in \mathbb{R}^{n \times d}$ by  
$$
\text{LayerNorm}(x;\gamma,\beta) = \gamma \circ \frac{x - \mu_x}{\sigma_x} + \beta,
$$
where $\mu_x,\sigma_x \in \mathbb{R}^n$ are the row-wise mean and standard deviation, and $\circ$ denotes the Hadamard (element-wise) product.  

Since $\mu_x$ and $\sigma_x$ are computed row-wise, they are invariant under column permutations, i.e., $\mu_{x\pi} = \mu_x$ and $\sigma_{x\pi} = \sigma_x$. Thus,  
$$
\text{LayerNorm}(x\pi;\gamma\pi,\beta\pi) 
= \gamma\pi \circ \frac{x\pi - \mu_x}{\sigma_x} + \beta\pi 
= \big(\gamma \circ \tfrac{x - \mu_x}{\sigma_x} + \beta\big)\pi 
= \text{LayerNorm}(x;\gamma,\beta)\pi.
$$

\textit{Proof of RMSNorm.}  
We aim to prove that  
$$
\text{RMSNorm}(x\pi;\gamma\pi) = \text{RMSNorm}(x;\gamma)\pi.
$$

The RMSNorm function is defined for $x \in \mathbb{R}^{n \times d}$ by  
$$
\text{RMSNorm}(x;\gamma) = \gamma \circ \frac{x}{\sqrt{\tfrac{1}{d}\sum_{j=1}^d x_j^2}},
$$
where the denominator is the root mean square computed over the feature dimension.  

Since $\sum_{j=1}^d (x\pi)_j^2 = \sum_{j=1}^d x_j^2$, the RMS value is permutation-invariant. Therefore,  
$$
\text{RMSNorm}(x\pi;\gamma\pi) 
= \gamma\pi \circ \frac{x\pi}{\sqrt{\tfrac{1}{d}\sum_{j=1}^d (x\pi)_j^2}} 
= \big(\gamma \circ \tfrac{x}{\sqrt{\tfrac{1}{d}\sum_{j=1}^d x_j^2}}\big)\pi 
= \text{RMSNorm}(x;\gamma)\pi.
$$

Thus, both LayerNorm and RMSNorm are permutation-equivariant.

\section{Experimental Setup Details}
\label{sec:app-exp}
For knowledge distillation, we used the Alpaca dataset and implemented training with PyTorch's Torchtune API. We conducted two separate distillation cases: (i) LLaMA-3.1-8B distilled into LLaMA-3.2-1B, and (ii) Qwen2.5-1.5B distilled into Qwen2.5-0.5B. Initial obfuscation and authorization of the teacher models were performed on an Intel SGX enclave, after which training and evaluation of DistilLock were carried out on an NVIDIA A100 (40GB) GPU. For the LLaMA case, we used the Adam optimizer with a learning rate of $1 \times 10^{-4}$, a KD ratio of 1.0, a batch size of 4, and gradient accumulation steps of 8. For the Qwen2 case, we used the Adam optimizer with a learning rate of $3 \times 10^{-4}$, a KD ratio of 0.5, a batch size of 8, and gradient accumulation steps of 8. Each experiment was repeated under three different random seeds. The total GPU cost for training and evaluation across both cases was approximately 70 GPU hours.

For adversarial training simulations, we modeled varying attacker capabilities. The white-box attack served as the best-case scenario for the adversary, assuming full access to both architecture and pre-trained weights. In this case, the attacker simply duplicated the original model and fine-tuned it on a separate labeled dataset, yielding the highest possible attack accuracy. The black-box attack represented the worst-case scenario, where the attacker was assumed to know only the architecture but not the weights. Hence, the surrogate model was initialized with random weights and trained from scratch on the attacker’s dataset.

For partial-defense baselines, we followed prior work to configure different levels of model parameter exposure~\cite{Li_2024}. In Serdab, the first Transformer block was protected inside the TEE and thus randomly initialized, while all remaining layers were accessible to the attacker. DarkneTZ instead protected the final Transformer block and classifier, which were randomly initialized, leaving earlier layers exposed. ShadowNet obfuscated all linear layers via matrix transformations; following prior work~\cite{zhang2023no}, we simulated the attacker by decoding these obfuscated layers to obtain surrogate initialization. For DistilLock, the surrogate model was initialized directly with obfuscated weights and then trained on the attacker’s labeled dataset. In all simulations, we sampled 1000 training and 1000 testing examples per task, following the assumption in prior work~\cite{Li_2024} that the attacker has access to only a small portion of the foundation model’s training data. The total GPU time consumed for these simulations was approximately 80 hours.

\end{document}